\documentclass[a4paper,12pt]{article}
\usepackage{latexsym}
\usepackage{exscale}
\usepackage{amsmath}
\usepackage{epsfig}
\usepackage{psfig}
\usepackage{graphicx}

\newcommand{\lsim}{{\;\raise0.3ex\hbox{$<$\kern-0.75em\raise-1.1ex\hbox{$\sim$}}
\;}}
\newcommand{\gsim}{{\;\raise0.3ex\hbox{$>$\kern-0.75em\raise-1.1ex\hbox{$\sim$}}
\;}}

\newcommand{\pr}[3]{{\it Phys.\ Rev.\ }{{\bf #1} {(#2)} {#3}}}

\newcommand{\prd}[3]{{\it  Phys.\ Rev.\ D\ }{{\bf #1} {(#2)} {#3}}}

\begin{document}
\baselineskip=20 pt
\setcounter{page}{1}
\thispagestyle{empty}
\hspace{11cm} HIP-2000-11/TH\\

\vspace{2 cm}
\centerline{\Large\bf Neutrinoless double beta decay }
\centerline{\Large\bf in four-neutrino models }
\vskip 1.5 cm
\centerline{{ Anna Kalliom\"aki} 
\footnote{E-mail address: amkallio@pcu.helsinki.fi} and { Jukka  Maalampi} 
 \footnote{E-mail address: maalampi@pcu.helsinki.fi}}
\vskip 0.8 cm
 \centerline{ \it{  Department of
Physics, Theoretical Physics Division}}
\centerline{ \it{ 
 University of Helsinki, Finland}}
\vskip 2 cm
\centerline{\bf ABSTRACT}\par
\vskip 0.5 cm
The most stringent constraint on the so-called effective electron neutrino mass from the present neutrinoless double beta decay experiments is  $\vert M_{ee}\vert < 0.2$ eV, while the planned next generation experiment GENIUS is anticipated to reach a considerably more stringent limit $\vert M_{ee}\vert < 0.001$ eV. We investigate the constraints these bounds set on  the neutrino masses and mixings of neutrinos in four-neutrino models where there exists a sterile neutrino along with the three ordinary neutrinos. We find that the GENIUS experiment would be sensitive to the electron neutrino masses down to the limit $m_{\nu_e}\lsim 0.024$ eV in such a scenario.


\newpage
\topskip 0  cm

There exist strong experimental indications that neutrinos have mass and that they mix.
The results of the Super-Kamiokande experiment \cite{atm} on  neutrinos produced in Earth's atmosphere by cosmic rays provide a convincing evidence of neutrino oscillations, and the results of the solar neutrino experiments \cite{solar} point to the same direction.  Results from  the laboratory experiment by the LSND collaboration \cite{LSND} may also be an indication of neutrino oscillations but they still await a future  confirmation, e.g. at the MiniBoone and I216 experiments 
\cite{doucet}, in particular so because measurements in KARMEN detector \cite{KARMEN} exclude most of  the region in the parameter space favoured by the LSND. 

Neutrino oscillation probabilities are determined by squared mass differences $\delta m_{ji}^2 = m_j^2-m_i^2$, where $m_i$ is the mass of the massive neutrino state $\nu_i$,  and the elements of the neutrino mixing matrix $U$ that connects the massive neutrino states $\nu_i$ and the flavour neutrino states $\nu_{\alpha}$ through the relation $\nu_{\alpha}=\sum_iU_{\alpha i}\nu_i$. For explaining the solar and atmospheric neutrino data two mass-squared difference scales, $\delta m_{\rm atm}^2\simeq  10^{-3}-10^{-2} \: {\rm eV}^2$ for atmospheric neutrinos and $\delta m_{\rm sun}^2\simeq  10^{-5} \: {\rm eV}^2$ or
$ \simeq 10^{-10} \: {\rm eV}^2$ for solar neutrinos, are needed, and the data can be described within a three-neutrino model. If the LSND result is not disregarded, a three-neutrino framework is not sufficient but one has to go beyond it and  consider a model with  four (or more) neutrinos as three different levels of mass-squared differences are needed, the third one being $\delta m_{\rm LSND}^2=0.3-2$ eV$^2$. In addition to the three known neutrino flavours $\nu_{e},\: \nu_{\mu}$ and $\nu_{\tau}$ one has to incorporate at least one light  neutrino $\nu_{s}$ that is sterile, i.e.~it lacks Standard Model (SM) gauge interactions \cite{four}. 

As was pointed out in ref. \cite{Walter}, the only four-neutrino mass patterns that are consistent with all oscillation data are those where there are two pairs of neutrinos with close masses, the "sun-pair" (mass states $\nu_0$ and $\nu_1$) and "atm-pair" (mass states $\nu_2$ and $\nu_3$) corresponding to the squared-mass differences $\delta m_{\rm atm}^2$ and 
$\delta m_{\rm sun}^2$, respectively, separated by a gap  corresponding to $\delta m_{\rm LSND}^2$ . 

Another oscillation constraint follows from cosmology. 
Depending on the value of $\delta m^2$, a large active-neutrino mixing could bring sterile neutrinos into thermal equilibrium
thereby increasing the effective number of light neutrinos \cite{original}, $N_{\nu}$, which would affect the big bang nucleosynthesis (BBN).    This leads to quite tight constraints on 
the active-sterile mixings \cite{original}--\cite{Shi}.

In the case neutrinos are Majorana particles \cite{valle}, which is quite likely  from the theoretical point of view, another useful source of experimental information on neutrino masses and mixings is provided by the neutrinoless double beta ($0\nu\beta\beta$) decay. The $0\nu\beta\beta$ decay is sensitive to the values of the neutrino masses themselves, in contrast with the oscillation data which depend only on the squared-mass differences. Also, it provides information on the Majorana CP-phases of neutrinos, to which the oscillation phenomena are practically insensitive. 
In the present work we shall study the constraints the $0\nu\beta\beta$ decay places on the neutrino masses $m_i$, the neutrino mixing matrix $U$ and the relative CP-phases of neutrinos in four-neutrino models consisting of three active neutrinos  $\nu_{e},\: \nu_{\mu}$ and $\nu_{\tau}$ and a sterile neutrino $\nu_{s}$ with the 2+2 mass pattern described above.

The quantity probed in the $0\nu\beta\beta$ decay experiments is the absolute value of the  effective mass
\begin{equation}
 M_{ee}=\sum_i m_iU_{ei}^2.
\label{Mee}\end{equation}
The most recent experimental upper bound for it from the Moscow-Heidelberg experiment is (at 90\% C.L.) \cite{Baudis}\begin{equation}
\vert M_{ee}\vert < 0.2\: {\rm eV}.
\label{Meelimit}\end{equation}
In turns out that this bound is in the case of four-neutrino models less restrictive than in the case of  the tree-neutrino models \cite{three}.
 A substantial strengthening of the present bound is, however, foreseen in the future. In the planned GENIUS experiment   one expects to reach after the first year of operation  the limit 0.01 eV  and eventually the limit \cite{GENIUS}
\begin{equation}
\vert M_{ee}\vert < 0.001\: {\rm eV}\;\;\; (\rm GENIUS).
\label{geniuslimit}\end{equation}
We will study the restrictions this  bound would set on the parameters of four-neutrino models, in particular on the mass of the electron neutrino $\nu_e$ and the mixing of $\nu_e$ with $\nu_{\mu}$ and $\nu_{\tau}$.
 
The constraint from the $0\nu\beta\beta$ decay is sensitive to the relative CP-phases $\eta_i$ of the mass states $\nu_i$. We will concentrate in the CP-conserving case where $\eta_i=\pm 1$. In this case one can write (\ref{Mee}) in the form \cite{BarKay}
\begin{equation}
 M_{ee}=\sum_i \eta_i  m_i\vert U_{ei}\vert^2.
\label{lMee}
\end{equation}
The most general mixing matrix $U$ that describes the connection between the flavour states $\nu_s,\nu_e,\nu_{\mu},\nu_{\tau}$  and  the  Majorana mass eigenstates $\nu_0,\nu_1,\nu_2,\nu_3$ can be
parametrized in terms of
$6$ rotation angles and $6$ phases as follows (see, e.g., ref. \cite{BarWhi}):
\begin{equation}
U = \left( \begin{array}{cccc}
c_{01}c_{02}c_{03} & c_{02}c_{03}s_{01}^*
& c_{03}s_{02}^* & s_{03}^* \\
\\
-c_{01}c_{02}s_{03}s_{13}^*
& -c_{02}s_{01}^*s_{03}s_{13}^*
& -s_{02}^*s_{03}s_{13}^*
& c_{03}s_{13}^*
\\
-c_{01}c_{13}s_{02}s_{12}^*
& -c_{13}s_{01}^*s_{02}s_{12}^*
& +c_{02}c_{13}s_{12}^*
&
\\
-c_{12}c_{13}s_{01}
& +c_{01}c_{12}c_{13}
&
&
\\ \\
-c_{01}c_{02}c_{13}s_{03}s_{23}^*
& -c_{02}c_{13}s_{01}^*s_{03}s_{23}^*
& -c_{13}s_{02}^*s_{03}s_{23}^*
& c_{03}c_{13}s_{23}^*
\\
+c_{01}s_{02}s_{12}^*s_{13}s_{23}^*
& +s_{01}^*s_{02}s_{12}^*s_{13}s_{23}^*
& -c_{02}s_{12}^*s_{13}s_{23}^*
&
\\
-c_{01}c_{12}c_{23}s_{02}
& -c_{12}c_{23}s_{01}^*s_{02}
& +c_{02}c_{12}c_{23}
&
\\
+c_{12}s_{01}s_{13}s_{23}^*
& -c_{01}c_{12}s_{13}s_{23}^*
&
&
\\ +c_{23}s_{01}s_{12} &
-c_{01}c_{23}s_{12}
&
&
\\ \\
-c_{01}c_{02}c_{13}c_{23}s_{03}
& -c_{02}c_{13}c_{23}s_{01}^*s_{03}
& -c_{13}c_{23}s_{02}^*s_{03}
& c_{03}c_{13}c_{23}
\\
+c_{01}c_{23}s_{02}s_{12}^*s_{13}
& +c_{23}s_{01}^*s_{02}s_{12}^*s_{13}
& -c_{02}c_{23}s_{12}^*s_{13}
&
\\
+c_{01}c_{12}s_{02}s_{23}
& +c_{12}s_{01}^*s_{02}s_{23}
& -c_{02}c_{12}s_{23}
&
\\
+c_{12}c_{23}s_{01}s_{13}
& -c_{01}c_{12}c_{23}s_{13}
&
&
\\
-s_{01}s_{12}s_{23}
& +c_{01}s_{12}s_{23}
&
&
\\ \\ \end{array} \right) \ ,
\label{Mixing}
\end{equation}
where $c_{jk} \equiv \cos\theta_{jk}$ and $s_{jk} \equiv \sin\theta_{jk} e^{i\delta_{jk}}$.
For our discussion  relevant is the second row which gives the composition of the electron neutrino in terms of the mass eigenstates neutrinos. It depends on several mixing parameters but fortunately simplifies considerably when the existing constraints from laboratory experiments and cosmology are taken into account. 

Assuming the 2+2 mass hierarchy among neutrino masses, indicating $\delta m^2_{02}\simeq \delta m^2_{03}\simeq \delta m^2_{12}\simeq \delta m^2_{13}\simeq \delta m^2_{LSND}$, the Bugey short baseline experiment \cite{bugey}, measuring the probability $P(\nu_e\to\nu_e)\simeq 1- A^{ee}\sin^2(L\delta m^2_{\rm LSND}/4E)$, yields the bound \cite{BarWhi} 
\begin{equation}
A^{ee}=4(\vert U_{e2}\vert^2+\vert U_{e3}\vert^2)(1-\vert U_{e2}\vert^2-\vert U_{e3}\vert^2)< 0.06,
\end{equation}
or
\begin{equation}
\vert c_{02}c_{13}s^*_{12}-s^*_{02}s_{03}s^*_{13}\vert^2 + \vert c_{03}s^*_{13}\vert^2< 0.016.
\label{Bugey0}
\end{equation}

The angles $\theta_{02}$ and $\theta_{03}$ are further constrained by the big bang nucleosynthesis (BBN) as large mixings would, as mentioned above, increase the number of  the effective degrees of freedom $N_{\nu}$ by bringing the sterile neutrino into thermal equilibrium.  According to a recent analysis  \cite{Shi}, the two-flavour $\nu_{\tau,\mu}\leftrightarrow \nu_s$ mixings should obey the constraint
(valid for $\vert\delta m^2\vert<2.5\cdot 10^{3} \;\; {\rm eV}^2$)
 \begin{equation}
\vert\delta m^2\vert\sin ^2\theta<7\cdot 10^{-5} \;\; {\rm eV}^2
\label{shi} 
\end{equation}
in order not to lead to a conflict with the BBN limit $N_{\nu}\lsim 3.2$ \cite{burles}. Since now $\delta m^2=\delta m_{\rm LSND}^2=0.3-2$ eV$^2$,  the mixing angle $\theta$ should be $\lsim 10^{-2}$, and  it is therefore conceivable to assume $s_{02}, s_{03}\ll 1$. With this approximation one derives from (\ref{Bugey0}) the bounds
\begin{eqnarray}
b&\equiv&\sin^2\theta_{12}+\sin^2\theta_{13}\lsim 0.016, \nonumber\\
 \vert b'\vert&\equiv&\vert\sin^2\theta_{12}-\sin^2\theta_{13}\vert\lsim 0.016,
\label{Bugey}
\end{eqnarray}
and  the following approximative composition of the electron neutrino in terms of the mass eigenstates:
\begin{equation}
\nu_e= -\sin\theta_{01}e^{i\delta_{01}}\nu_0-\cos\theta_{01}\nu_1+\sin\theta_{12}e^{i\delta_{12}}\nu_2+\sin\theta_{13}e^{i\delta_{13}}\nu_3.
\label{nuecomp}
\end{equation}

Let us note that the cosmological argument  used above remains still somewhat controversial. Some analysis allow for  the number of  the relativistic degrees of freedom, instead of the bound $N_{\nu}\lsim 3.2$ quoted above, values close to four \cite{Olive}, in which case BBN would not place any constraint on the angles $s_{02}, s_{03}$. The difference is due to conflicting results on the primordial deuterium abundance. 
Furthermore, the BBN constraint depends on the lepton asymmetry in the early universe, which might considerably weaken the bounds \cite{fvA} (in the case the predominantly sterile mass eigenstate is lighter than its active mixing partner). Anyhow, we will assume in the following that the sterile neutrino mixes considerably only with the electron neutrino and consequently that the approximations (\ref{Bugey}) and (\ref{nuecomp}) are valid.

In this approximation  the solar neutrino oscillation is described entirely in terms of the  angle $\theta_{01}$ associated with the $\nu_e\leftrightarrow\nu_s$ mixing. According to an overall analysis of all solar neutrino data, only the small angle matter solution (SAMSW) with $2\cdot10^{-3}\lsim \sin^{2}2\theta_{01}\lsim 10^{-2}$ leads to a reasonable fit in the active-sterile mixing case \cite{BahSmi}.

There are two mass patterns consistent with the oscillation data, differing in as to which one of the pairs, the sun-pair or the atm-pair, is lighter. We will call Model A the pattern where the sun-pair is the light one and the atm-pair is the heavy one, and the case where the mass order is the opposite will be called Model B. 
The mass scale of the lighter neutrino pair is denoted by $m$, and the mass gap between the pairs is denoted by $\Delta m$, so that the mass scale of the heavier pair is $m+\Delta m$. We will neglect the mass differencies inside the two pairs by assuming they to be small as compared with the mass gap $\Delta m$.

Studies of the anisotropies of the cosmic microwave background radiation and the large scale structure of the universe provide information on the sum of neutrino masses, $\sum_i m_{\nu_i}\simeq 4m+2\Delta m$. The contribution of neutrinos on the dark matter content of the Universe is in units of the critical density given by $\Omega_{\nu}=h^{-2}\sum_i m_{\nu_i}/94$ eV, where $h$ is the dimensionless Hubble constant ($H_0=100\:h \;{\rm km\;s}^{-1}\;{\rm Mpc}^{-1}$). In the previously popular hot+cold dark matter scenario \cite{hotcold}   the neutrino contribution to the energy density of the universe is assumed to be $\Omega_{\nu}\simeq 0.2$, but the more recent observations, indicating the existence of a large cosmological constant, imply that this should be taken as a conservative upper limit only \cite{kayser}. With the  recent measured value of the Hubble constant \cite{hubble}, $h\simeq 0.71$, this leads to the upper limit $\sum_i m_{\nu_i}\lsim 10$ eV. Recently it has been claimed that this limit improves to 
\begin{equation}
\sum_i m_{\nu_i}\lsim 5.5 \;\; {\rm eV}
\label{aniso}
\end{equation} 
 when the information from the Ly$\alpha$ forest in quasar spectra is taken into account \cite{croft}. 

The range of  the squared mass difference, $\delta m_{\rm LSND}^2=0.3-2$ eV$^2$, indicated by the LSND data and allowed by the Bugey and the other short baseline experiments  yields the following upper and lower bounds for the mass difference $\Delta m$ as a function of the mass $m$:
\begin{equation}
-m+\sqrt{m^2+0.3\;{\rm eV}^2}\;\lsim\;\Delta m\;\lsim\;-m+\sqrt{m^2+2\;{\rm eV}^2}.
\label{Dmrange}
\end{equation}
Combining this with the cosmological bound (\ref{aniso}) leads to the following bounds on $m$ and $\Delta m$:  
\begin{eqnarray}
m&\lsim& 1.3 \;\; {\rm eV},\nonumber\\
\Delta m&\simeq&\frac{\delta m_{\rm LSND}}{\sum_i m_i/2}\;\geq \; 0.1  \;\; {\rm eV}.
\label{Dmcosmo}
\end{eqnarray}
In the extreme case of  $m=1.3$ eV and  $\Delta m=0.1$ eV the mass spectrum is quite degenerate with $m_0\simeq m_1\simeq 1.3$ eV and $m_2\simeq m_3\simeq 1.4$ eV (Model A) or $m_0\simeq m_1\simeq 1.4$ eV and $m_2\simeq m_3\simeq 1.3$ eV (Model B). At the other extreme of $m=0$, i.e.~when the lighter neutrino pair is massless, the heavier neutrino pair would have its mass in the range 0.5 eV to 1.4 eV.

Let us proceed to examine the consequences of the constraints (\ref{Meelimit}) and (\ref{geniuslimit}) on  the neutrino masses and mixing angles $\theta_{01},\; \theta_{12}$ and $\theta_{13} $ for different CP-phase patterns $(\eta_0,\eta_1,\eta_2,\eta_3)$ in the two models A and B.  
 
{\bf Model A.}
In this model the neutrino pair ($\nu_0,\nu_1$) responsible for the solar anomaly  is lighter than the pair ($\nu_2,\nu_3$) responsible for the atmospheric neutrino anomaly, i.e.~the mass spectrum is of the form
\begin{eqnarray} 
&&m_0\simeq m_1=m\nonumber\\ 
&&m_2\simeq m_3=m+\Delta m.
\end{eqnarray}
The $0\nu\beta\beta$ condition takes in this case the form
\begin{equation}
|m(\eta_0\sin^2\theta_{01}+\eta_1\cos^2\theta_{01})+(m+\Delta m)(\eta_2\sin^2\theta_{12}+\eta_3\sin^2\theta_{13})|<a , 
\label{eka}
\end{equation}
where for the present experimental bound   $a=0.2$ eV and for the anticipated future bound of the GENIUS experiment $a=0.001$ eV. In order to find out the consequences of this condition one should examine separately the following different patterns of the relative CP-phases $\eta=(\eta_0,\eta_1,\eta_2,\eta_3)$:
\begin{eqnarray}
&{\rm  I, II}&\eta=(1,1,\pm 1,\pm 1),\nonumber\\
&{\rm III,  IV}&\eta=(1,1,\pm 1,\mp 1),\nonumber\\
&{\rm  V, VI}&\eta=(1,-1,\pm 1,\pm 1),\nonumber\\
&{\rm  VII, VIII}&\eta=(1,-1,\pm 1,\mp 1).
\label{etacombinations}
\end{eqnarray}

In the first four cases (I-IV), where $\eta_0=\eta_1$, the $0\nu\beta\beta$ condition is not affected by the solar neutrino mixing angle $\theta_{01}$. 

In Cases I and II the $0\nu\beta\beta$ condition (\ref {eka}) becomes
\begin{equation}
 \vert m\pm (m+\Delta m) b\vert < a
\label{ekaI_II}\end{equation}
where the plus-sign corresponds to Case I and the minus-sign to Case II, and $b=\sin^2\theta_{12}+\sin^2\theta_{13}$, as defined in eq. (\ref{Bugey}). 

{\it Case I}. The LSND and other short baseline experiments constrain the value of the mass gap $\Delta m$ between the two neutrino pairs within the range given in eq. (\ref{Dmrange}). Combining this with the cosmological  lower bound (\ref{Dmcosmo}) yields for $\Delta m$ the allowed range from 0.1 eV to 1.4 eV. One then concludes from (\ref{ekaI_II}) that with the present experimental sensitivity the $0\nu\beta\beta$ result does not set any constraint on the mixing angles $\theta_{12},\theta_{13}$ as $b\Delta m\ll a$. The experimental constraint  implies just an upper limit for the mass scale $m$,  $m\lsim (a-b\Delta m)/(1+b)\simeq a = 0.2$ eV.

 The situation would be different in the case of the anticipated GENIUS bound $a=0.001$ eV. The condition (\ref{ekaI_II}) to be satisfied one must now have
$ b\le a/\sqrt{0.3}\;{\rm eV}\simeq 2a/{\rm eV}=0.002$ , which would mean  tightening of  the present bound (\ref{Bugey}) from the Bugey reactor experiment by one order of magnitude. Also the allowed values of $m$ are relatively small, $m\lsim a=0.001$ eV.  

\begin{figure}[h]\label{kolkkikuva}
\centering
\vspace{10mm}
\begin{tabular}{@{\hspace*{0.5cm}}c}
\mbox{\epsfig{file=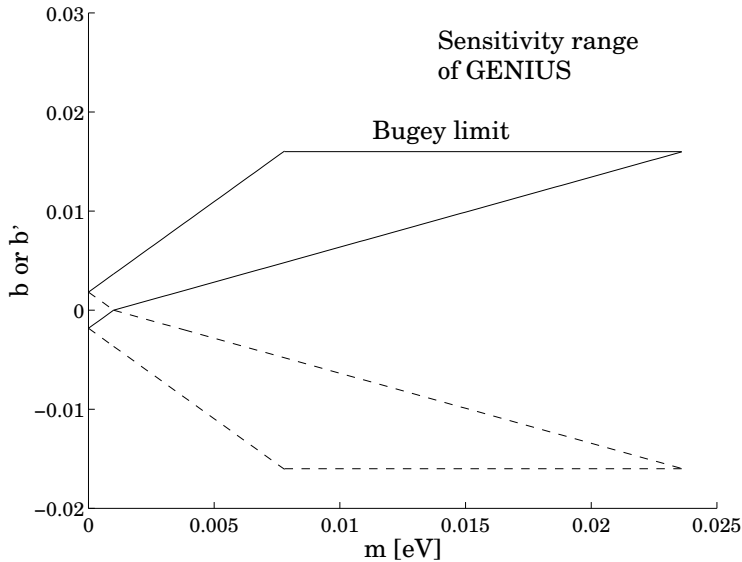,width=10cm,angle=0}}
\end{tabular}
\vspace{1cm}
\caption{\small The sensitivity range reached by the planned GENIUS experiment in the mass parameter $m$ and the mixing angle variables $b=\sin^2\theta_{12}+\sin^2\theta_{13}$ and $b'=\sin^2\theta_{12}-\sin^2\theta_{13}$ in Model A. The upper and/or lower bounds of $\pm 0.016$ for $b$ and $b'$ are obtained from Bugey experiment \cite{bugey}. The area bounded by the dashed line illustrates Cases I ($b$ as vertical axis) and III ($b'$ as vertical axis)  (see eq.(\ref{etacombinations})) whereas the area bounded by solid line illustrates the sensitivity range in Cases II ($b$) and IV ($b'$). Cases V-VIII are related to Cases I-IV in a way described in the text.}    
\end{figure} 

{\it Case II}. The Case II does not differ from Case I in any essential way as far as the present $0\nu\beta\beta$ bound is concerned. With the GENIUS bound  the upper limit of the mixing angles $\theta_{12},\theta_{13}$ is as  stringest  $b \lsim a/\sqrt{0.3}\;{\rm eV}\simeq 2a/{\rm eV}= 0.002$ obtained for $m=0$. 
The situation is different compared with Case I in that now the condition (\ref{ekaI_II}) imposes also a non-zero lower limit on $b$ when $m \gsim a $, otherwise there would not be sufficient cancellations among the varios terms to make the condition (\ref{eka}) to be valid. The quantity $b$ would be constrained into the range $(m-a)/\sqrt{m^2 + 2 \;{\rm eV^2}}\lsim b\lsim(m+a)/\sqrt{m^2 + 0.3 \;{\rm eV^2}}$, where the upper limit is overtaken by the upper limit $b\lsim 0.016$ from the short baseline experiments, when $m\gsim 0.008$ eV. The sensitivity range of the GENIUS experiment in the $(m,b)$ parameter plane is presented in Fig.~1 (solid line).
As can be read from the figure,  the GENIUS experiment would in this case be sensitive to the values of $m$ down to 0.024 eV.

{\it Cases III and IV}. In Cases III and IV the $0\nu\beta\beta$ condition (\ref {eka}) becomes
\begin{equation}
 \vert m \pm (m+\Delta m) b'\vert < a,
\label{ekaIII_IV}\end{equation}
where the plus-sign corresponds to Case III and the minus-sign to Case IV, and  $b'=\sin^2\theta_{12}-\sin^2\theta_{13}$, as defined in (\ref{Bugey}). Just like in Cases I and II, the present $0\nu\beta\beta$ bound does not probe the mixing angles $\theta_{12}$ and $\theta_{13}$ but sets a constraint merely on the mass $m$. Again, the GENIUS experiment would be sensitive to the values of these angles. In Case III the condition (\ref{ekaIII_IV}) restricts the allowed values of $b'=\sin^2\theta_{12}-\sin^2\theta_{13}$, whose value is by other constraints limited to $-0.016\lsim b'\lsim 0.016$, so that for example for $m=0$ only the values $|b'|\le a/\sqrt{0.3}\;{\rm eV} \simeq 2a/\textrm{eV}=0.002\,$ are allowed.  The condition (\ref{ekaIII_IV}) is never satisfied  for the mass values $m\gsim  0.024\,$ eV. This would be the sensitivity limit of the GENIUS on the electron neutrino mass also in this case. The allowed values of $b'$ are negative, $(-m-a)/\sqrt{m^2 + 0.3\;{\rm eV^2}}\lsim b'\lsim (-m+a)/\sqrt{m^2 + 2 \;{\rm eV^2}}$, except that for $m<a$ small positive values are also possible. The GENIUS sensitivity range for this case is presented in Fig.~1 with the dashed line.
Case IV is a mirror image of Case III so that again for $m=0$ all values $|b'|\lsim a/\sqrt{0.3}$ eV are possible, but now $b'$ can have mainly positive values: $(m-a)\sqrt{m^2 + 2 \;{\rm eV^2}}\lsim b'\lsim(m+a)/\sqrt{m^2 + 0.3 \;{\rm eV^2}}$ (the solid line in Fig.~1). Note that this is almost the same region that was obtained in Case II for the quantity $b$. The  limit that the GENIUS experiment could reach for the mass $m$ is again 0.024 eV. 

{\it Cases V and VI}. In these cases the $0\nu\beta\beta$ condition depends also on the solar mixing angle $\theta_{01}$: 
\begin{equation} 
|m (\sin^2\theta_{01}-\cos^2\theta_{01})\pm (m+\Delta m)b|<a\,,
\label{ekaV_VI}\end{equation}
where the plus-sign corresponds to Case V and the minus-sign to Case VI. If only the small mixing angle (SMA) MSW solution is being considered, one can use the approximative value $\sin^2\theta_{01}-\cos^2\theta_{01}\simeq \pm 1$, where the sign depends on whether $\theta_{01}$ is close to $0$ or $\pi/2$. If one assumes that angle $\theta_{01}\simeq0,$ condition (\ref{ekaV_VI}) becomes 
\begin{equation}
|-m\pm (m+\Delta m)b|\lsim a\,.\label{tokaIII_IV}
\end{equation}
As is apparent from eqs.~(\ref{ekaI_II}) and (\ref{tokaIII_IV}), the $0\nu\beta\beta$ constraints in Cases II and V are in practical terms the same, and  no separate treatment is needed for Case V. The same applies to Cases I and VI. On the other hand, if the angle $\theta_{01}$ is close to $\pi/2$, condition (\ref{tokaIII_IV}) must be replaced with 
\begin{equation}
|m\pm (m+\Delta m)b|\lsim a\,.\label{kolmasIII_IV}
\end{equation}
and clearly Cases I and V as well as Cases II and VI correspond to each other as far as the $0\nu\beta\beta$ constraint is concerned.  

It is noteworthy that the values of $\theta_{01}$ close to 0 and close to $\pi/2$ correspond here two physically distinct situations while in oscillation experiments, where transition probabilities depend on $\theta_{01}$ through $\sin^22\theta_{01}$, do not make any difference between the ranges $0\lsim \theta_{01}\lsim \pi/4 $ and $\pi/4\lsim \theta_{01}\lsim\pi/2$. 

{\it Cases VII and VIII}. These cases are practically identical to Cases III and IV. If $\theta_{01}\simeq 0,$ the $0\nu\beta\beta$ condition~(\ref{eka}) takes the form
\begin{equation}
|-m\pm(m+\Delta m)b'|<a,
\label{ekaVII_VIII}
\end{equation} 
where  the plus-sign corresponds to Case VII and minus-sign to Case VIII. Comparison between eqs.~(\ref{ekaIII_IV}) and (\ref{ekaVII_VIII}) shows that in Cases IV and VII as well as in Cases III and VIII the $0\nu\beta\beta$ condition is practically the same. If $\theta_{01}\simeq\pi/2,$ the same applies to Cases III, VII and IV, VIII.  
     
{\bf Model B.}
In this model the mass spectrum is of the form 
\begin{eqnarray} 
&&m_0\simeq m_1=m+\Delta m\nonumber\\ 
&&m_2\simeq m_3=m.
\end{eqnarray}

Again we assume that mixing of the neutrino pair $(\nu_0,\nu_1)$ is mainly responsible for the solar neutrino anomaly and mixing of the pair $(\nu_2,\nu_3)$ for the atmospheric neutrino anomaly. The $0\nu\beta\beta$ condition becomes now
\begin{equation}
|(m+\Delta m)(\eta_0\sin^2\theta_{01}+\eta_1\cos^2\theta_{01})+m(\eta_2\sin^2\theta_{12}+\eta_3\sin^2\theta_{13})|<a .   \label{modelB}
\end{equation}
If LSND results are taken into account and one assumes that $0.3\,\textrm{eV}^2\lsim m_1^2-m_3^2\lsim 2\,\textrm{eV}^2$, which yields $m_1\gsim 0.5\,\textrm{eV}$, it is easy to see that eq.~(\ref{modelB}) is not realized for any value of $m$ \cite{Bilenky}. This is because the absolute values of $b$ and $b'$, defined in eq.~(\ref{Bugey}), are much less than $|\eta_0\sin^2\theta_{01}+\eta_1\cos^2\theta_{01}|\simeq 1$, so that eq.~(\ref{modelB}) can be written approximately as \begin{equation}
|m+\Delta m|\lsim a\,,\label{likiarvo}
\end{equation}
which is untrue because $|m+\Delta m|=m_1$ and even the present value of $a$ is less than the lower limit $0.5\,$eV of $m_1$. If we, on the other hand, disregard the LNSD results and  let $\Delta m$ be as low as $0.1\,$eV, with present value of $a$ we get some restrictions on the parameter values from eq.~(\ref{modelB}). Eight different $\eta$ combinations (\ref{etacombinations}) are in principle possible also in this case but because (as was already mentioned) the term proportional to $ b $ or $ b' $ in eq.~(\ref{modelB}) is much smaller than the term proportional to angle $\theta_{01}$, it can safely be neglected. So, the approximative  equation (\ref{likiarvo}) is valid in every case, not depending on the choices of $\eta_{0}$ and $\eta_{1}$ or whether $\theta_{01}$ is $\simeq 0$ or $\pi/2$. The allowed values for $m$ are in the range $m \lsim a-\Delta m$.   

\bigskip

We can summarize our results as follows. In the model A, i.e.~when the neutrino pair (with mass $m$) responsible on the solar neutrino deficit is lighter than the neutrino pair (with mass $m+\Delta m$) responsible on the atmospheric neutrino anomaly,   the $0\nu\beta\beta$ constraint set by the future GENIUS experiment, $M_{ee}\lsim 0.001$ eV,  induces an upper limit for mass $m$ as a function of  the quantity $b=\sin^2\theta_{12}+\sin^2\theta_{13}$ or the quantity $b'=\sin^2\theta_{12}-\sin^2\theta_{13}$ depending on the pattern of the relative CP numbers of the neutrinos. There exists eight  possible patterns of the relative CP numbers, Cases I to VIII, defined in eq.~(\ref{etacombinations}).  In Case I, where all relative CP numbers are equal, and no cancellations between the contributions of different neutrinos hence occurs, the upper bound for $m$ is the stringest, $m\lsim 0.001$ eV. For all the others Cases the absolute mass bound is about an order of magnitude less stringent, $m\lsim 0.024$ eV. In Case I  the value of $b$ must be small, $b\lsim a$.  In the other Cases  the restriction on  $b$ or $b'$ depend on the value of $m$ (see Fig.~1), and both a lower and an upper limit are obtained. For $m\lsim 0.08 $ eV the foreseen upper limit for $b$ or $\vert b'\vert$ is more stringent that the limit of the Bugey experiment. There is a crucial difference between the different relative CP number patterns as comes to the ensuing limitations on the mixing angles  $\theta_{12}$ and $\theta_{13}$, as a bound on $b$ implies a bound on the both angles whereas a bound on $b'$ implies only a bound on the difference of the angles, not on each of the angles separately. Also, the smaller is the mass $m$, the smaller  or the more degenerate are the angles  $\theta_{12}$ and $\theta_{13}$, depending on whether the obtained bound is on $b$ or $b'$, respectively.

The Model B, i.e.~when the neutrino pair responsible on the solar neutrino deficit is heavier than the neutrino pair  responsible on the atmospheric neutrino anomaly, is in practical terms ruled out already by the present limit of $M_{ee}\lsim 0.2$ eV. 

\bigskip

This work has been supported by the Academy of Finland under the project no. 40677.

\end{document}